# Characterization of nanometer-sized, mechanically exfoliated graphene on the H-passivated Si(100) surface using scanning tunnelling microscopy

Kyle A. Ritter[1,2] and Joseph W. Lyding[1,3]

[1] Beckman Institute for Advanced Science and Technology, University of Illinois, 405 North Mathews Avenue, Urbana, IL 61801-2325, USA

[2] Department of Materials Science and Engineering, University of Illinois, 1304 West Green Street, Urbana, IL 61801-2920, USA

[3] Department of Electrical and Computer Engineering, University of Illinois, 1406 West Green Street, Urbana, IL 61801-2918, USA

E-mail: ritter1@uiuc.edu

**Abstract**

We have developed a method for depositing graphene monolayers and bilayers with minimum lateral dimensions of 2-10 nm by the mechanical exfoliation of graphite onto the Si(100)–2×1:H surface. Room temperature, ultra-high vacuum (UHV) tunnelling spectroscopy measurements of nanometer-sized single-layer graphene reveal a size dependent energy gap ranging from 0.1-1 eV. Furthermore, the number of graphene layers can be directly determined from scanning tunnelling microscopy (STM) topographic contours. This atomistic study provides an experimental basis for probing the electronic structure of nanometer-sized graphene which can assist the development of graphene-based nanoelectronics.

*UHV-STM characterization of nanometer-sized graphene on H-Si(100)*

**1. Introduction**

The isolation of graphene, a single atom thick sheet of graphite, by mechanical exfoliation has enabled researchers to probe its intrinsic electronic properties which include a mobility value ten times greater than Si and submicron ballistic transport [1]. The ability to tailor the electronic structure of graphene using conventional lithographic methods makes graphene a desirable material for future electronic devices. In particular, the lithographic patterning of graphene circumvents a number of critical large-scale integration issues, including accurate positioning, orientational control, and low-resistance contacts, associated with other electronic nanomaterials such as carbon nanotubes and semiconducting nanowires. Researchers have tailored the electronic structure of graphene monolayers by utilizing electron beam lithography to fabricate graphene with lateral dimensions as small as 10 nm, enabling a graphene-based single electron transistor [1] and semiconducting graphene nanoribbons (GNRs) [2,3]. The advancement of graphene-based nanoelectronics [4,5,6] requires a detailed understanding of the atomic and electronic structure of nanometer-sized graphene. Due to its atomic-scale imaging and spectroscopic capabilities, scanning tunnelling microscopy (STM) provides a means for characterizing graphene with nanometer-scale lateral dimensions. Several STM studies of graphene have emerged which include mechanically exfoliated graphene on $SiO_2$ [7,8], epitaxial graphene on SiC [9,10,11,12], and graphene synthesized by chemical vapor deposition of hydrocarbons on noble metal surfaces [13,14,15,16]. However, previous STM studies have focused on graphene layers with a minimum lateral size of a few microns, whereas dimensions of tens of nanometers are required to induce a semiconducting gap due to the lateral confinement of charge carriers [17,18,19].

We have extended the dry contact transfer (DCT) [20] technique to isolate predominantly single layers of graphene with minimum lateral dimensions ranging from 2-10 nm on the inert Si(100)–2×1:H surface. The nanometer-sized graphene pieces are formed spontaneously during the mechanical exfoliation of graphite. Previous experiments utilizing mechanically exfoliated graphene [1] have been unable to locate graphene monolayers with lateral sizes beneath the diffraction limit of visible light due to the necessity of using optical microscopy to identify the single-layer graphene. Furthermore, the number of graphene layers can be directly determined from the STM topograph by measuring the height of the graphene. STM studies of mechanically exfoliated gra-phene on $SiO_2$ required additional instrumentation, including atomic force microscopy [7] and Raman spectroscopy [8], to determine the number of graphene layers due to the electrically insulating nature of the substrate. We have also measured a size-dependent energy gap for graphene monolayers with dimensions ≤ 10 nm using scanning tunnelling spectroscopy (STS). Since room-temperature tunnelling spectroscopy is limited to an accuracy of 100 meV due to thermal broadening [21], the largest graphene piece which should possess an energy gap that can be detected with STS at room temperature would be ≈40 nm, which has been predicted to exhibit a maximum energy gap of 100 meV for armchair GNRs [19]. Previous tunnelling spectroscopy studies of graphene on SiC [9,10], Ir [15], and Ru [16] surfaces have reported a finite density of states near the Fermi level since the graphene layers are continuous over macroscopic regions of the surface and bandgaps induced from spatial confinement are not expected. Although gap-like features of 100 meV have been observed for graphene monolayers and bilayers on SiC, the authors attributed the observed energy gap to a combination of charging and band bending effects or inelastic coupling to surface excitations [12].

**2. Experimental Details**

Our experiments were conducted using a homebuilt room temperature ultra-high vacuum scanning tunnelling microscope (UHV-STM) at a base pressure of $5 \times 10^{-11}$ Torr [22]. In our experimental setup, the bias voltage is applied to the sample and the tip is grounded through a current preamplifier. STS consists of acquiring a tunnelling current-voltage (*I-V*) spectrum by momentarily disabling the feedback at predefined points in the topographic image and sweeping the voltage over a specified range while recording the current. Topographic images were collected using current setpoint values ranging from 50-100 pA. To increase the tunnelling current at the band edges in the STS data, the tip-sample spacing was decreased relative to the topographic image current setpoint by increasing the current to 1 nA before collecting the tunnelling *I-V* spectra.

Experiments were performed on degenerately As-doped n-type Si(100) substrates with a resistivity < 0.005 Ω cm using electrochemically etched tungsten tips. The Si substrate was hydrogen-passivated by leaking $H_2$ into the UHV chamber where atomic H was created by cracking the $H_2$ molecules on a hot (1500 °C) tungsten filament. Graphene was deposited onto the Si(100) – 2×1:H surface using an *in-situ* DCT method [20]. Graphene was mechanically exfoliated by rubbing a highly oriented pyrolytic graphite crystal [23] against an alumina crucible which produced a fine, black powder. In the DCT method, braided fiberglass is loaded with exfoliated graphite powder and the applicator is subsequently heated in UHV to remove physisorbed ambient molecules. The applicator is then manipulated into gentle contact with the sample surface, which yields predominantly single layers of graphene. For the 74 graphene pieces identified with the STM, the distribution consisted of 82% monolayers, 11% bilayers, and 7% three to five layers thick graphene. The minimum and maximum lateral dimensions of the graphene were 6 ± 4 nm and 18 ± 13 nm, respectively, and the thicker graphene pieces exhibited larger average dimensions than the graphene monolayers.

## 3. Results and Discussion

Figure 1(a) shows a 100 x 100 $nm^2$ STM image after stamping the H-passivated Si(100)–2×1 surface with the graphene-loaded DCT applicator. In the upper right corner of the image, there is a feature exhibiting an average lateral dimension of ≈6 nm. The nanostructure from figure 1(a) is shown at higher resolution in figure 1(b). The figure 1(b) inset presents the spatial derivative of a subsequent STM topograph of the feature which depicts several parallel lines spaced by 2.5 ± 0.4 Å, consistent with the 2.5 Å spacing between nearest neighbor carbon hexagons [24]. A line contour taken from the topographic image of the figure 1(b) inset derivative image is displayed in figure 1(c), highlighting the 2.5 Å modulation of the feature. The white line in the figure 1(b) inset delineates the position of the contour in figure 1 (c). Figure 1(d) shows a cross section taken from the topographic image along the gray line in figure 1(b); the feature displays a ≈3.1 Å height relative to the H-passivated Si dimers, comparable to the expected sum of the thickness of single-layer graphene and the Van der Waals bonding distance. Based on the atomic-scale structure observed in the inset of figure 1(b) and the apparent height from the STM topograph, the feature presented in figure 1 is consistent with a graphene monolayer. Interestingly, we observe a ≈1 Å protrusion near the center of the graphene monolayer piece in figure 1(b) which is apparent in the STM height profile displayed in figure 1(d). From the figure 1(b) inset, we do not observe local distortion of the graphene lattice near the protrusion suggesting that the increased topography is not due to a defect intrinsic to the graphene lattice as was observed recently in low temperature STM studies of epitaxial graphene on SiC [10]. Due to the presence of Si dangling bonds on the nearby substrate, we believe this localized protrusion in the graphene topography is most likely due to an underlying, unpassivated Si atom, simi-lar to SiC adatoms at the graphene/substrate interface which appear in the STM topograph of graphene monolayers [10]. Based on previous experiments using single-walled carbon nanotubes supported on the Si(100)–2×1:H surface [25,26], future experiments integrating graphene monolayers with patterns of STM-depassivated Si could be used to verify whether underlying Si dangling bonds induce localized protrusions in the graphene topography.

In addition to graphene monolayers, we have also observed nanometer-sized graphene bilayers. Figure 2(a) highlights an image where simultaneous atomic-scale resolution is obtained for both the 10 x 15 nm$^2$ graphene bilayer and the substrate. A line contour from the STM topograph along the gray line in figure 2(a) is shown in figure 2(b). The contour intersects both monolayer and bilayer regions of the graphene feature. The measured graphene monolayer height is ≈ 2.8 Å, consistent with the measured height of the graphene monolayers from figure 1. The bilayer has a ≈ 5.7 Å height relative to the Si dimers and the distance between the two graphene layers is 2.9 Å which is somewhat smaller than the known 3.35 Å interlayer spacing [24] in bulk graphite. Figure 2(c) shows an enlarged portion of the topographic spatial derivative of figure 2(a) taken from the interior of the bilayer. We observe features in the topograph of the graphene bilayer that are consistent with STM topographs of graphite where three of the carbon atoms in each hexagon are imaged as protrusions due to the asymmetric stacking between the graphene layers [27].

Tunnelling spectroscopy data were also collected for the graphene monolayer in figure 1, with the current-voltage (*I-V*) spectra for the graphene and H-passivated Si plotted on a semilogarithmic scale in figure 3. The spatial locations of the *I-V* spectra recorded on the graphene and Si are indicated by the blue and red circles, respectively, in the figure 3 inset. For the graphene tunnelling spectrum, we measure a bandgap of 0.3 eV. For the spectra recorded on the substrate, we measure an energy gap of 1.4 eV, 0.3 eV higher than the 1.1 eV Si bulk bandgap. The control spectra are broadened by 30% due to a voltage drop across the Si resulting from dopant outdiffusion [28]. If we assume 30% broadening of the graphene energy gap due to band bending, the 0.06 eV error is negligible compared to the 0.1 eV uncertainty due to thermal broadening [21]. The 0.06 eV represents the maximum broadening of the graphene spectra due to substrate bandbending since the current density likely decreases after exiting the graphene which would decrease the voltage dropped across the substrate.

*UHV-STM characterization of nanometer-sized graphene on H-Si(100)*

Previous STS studies of SWNTs on highly doped Si(100) – 2x1:H surfaces [20,29] suggest that the electronic structure of the graphene can be accurately probed with STS and the graphene electronic properties are not perturbed by H-passivated Si. Tunnelling spectroscopy studies of semiconducting SWNTs on the Si(100) – 2x1:H surface [20,29] observed bandgap values consistent with the theoretically predicted values for ≈1 nm diameter SWNTs implying that the bandgap of the Si substrate does not influence the accuracy of the measured energy gap values. Due to the chemical similarities between SWNTs and graphene, and that first-principles theoretical studies of SWNTs on H-passivated Si(100) conclude that SWNTs are physisorbed [29], we can infer that the graphene interacts weakly with the H-Si surface. Although *I-V* curves were not collected directly over the previously discussed ≈ 1 Å localized protrusion in the middle of the graphene, a suspected effect of the underlying unpassivated Si, tunnelling spectroscopy will be used in future studies to understand the effect of clean Si on the graphene electronic structure.

Due to the ≈ 6 nm dimensions of the graphene, we propose that the 0.3 eV bandgap of the graphene is due to the quantum confinement of carriers. Graphene nanoribbons (GNRs), strips of graphene monolayers with nanometer widths and lengths up to several microns, have been theoretically predicted [17,18,19] and experimentally observed [2,3] to yield a bandgap which decreases inversely with length. In addition to width, the GNR bandgap depends on crystallographic orientation and theoretical studies suggest that GNRs with widths of 6-8 nm can display gaps of 0.04-0.4 eV depending on their edge structure [17,18,19]. Unlike GNRs, the graphene monolayers that we observe experimentally are constrained in all directions and the crystallographic orientation along the edges is generally not well defined. However, we do observe that the energy gaps on the nanometer-sized graphene monolayers decreases inversely with its lateral dimensions, as shown in figure 4, which further supports our interpretation that the observed graphene bandgaps result from finite-size effects.

Figures 4(a) and (b) depict the STM topograph and spatial derivatives images, respectively, of two graphene monolayers where heights of 3-3.5 Å were measured from the center of both pieces relative to the Si dimers. The upper piece has lateral dimensions of $1.8 \pm 0.4 \times 3.2 \pm 0.7$ nm$^2$ while the lower piece is $4.7 \pm 1.0 \times 8.3 \pm 1.0$ nm$^2$. Figures 4(c) and (d) contain spectra maps, where the log(*I*)-*V* curves are plotted as a function of position with 4 Å spatial resolution along the red lines in figure 4(a), which elucidate local variations in elec-tronic structure of the graphene and the proximal Si surface. The spectra collected over the graphene are bounded by the white dotted lines in figures 4(c) and (d). The measured substrate valence and conduction band edges are labeled in figures 4(c) and (d) and the 1.2 eV energy gap closely agrees with the expected 1.1 eV bulk Si bandgap. The most prominent feature observed in the spectra maps is that the graphene energy gap scales inversely with the lateral size of the graphene monolayers, consistent with theoretical predictions [17,18,19]. The lower, 5 nm wide graphene monolayer displays a spatially varying energy gap of 0.1 - 0.3 eV, while the upper, 2 nm wide piece ranges from 0.4 – 1.0 eV. Although the edge functionalization and crystallographic orientation are unknown for the graphene monolayers in figure 4, recent studies of finite-length, armchair GNRs with lateral dimensions comparable to the 1.8 x 3.2 nm$^2$ graphene monolayer yield similar band gap values to those measured experimentally. Shemella *et. al* [30] have reported density functional calculations on a 1.3 x 2.6 nm$^2$ armchair GNR with hydrogen passivated edges which exhibits a 0.5 eV energy gap which is within the 0.4-1.0 eV range of measured values. We anticipate that future theoretical calculations which consider the detailed geometry of the graphene observed experimentally will provide additional insight into the energy gap values obtained from STS measurements.

The variation in the experimental graphene energy gap values likely results from the unique atomic structure at the edges of the graphene layers, which depends on the crystallographic orientation and functionalization of the carbon atoms along the edge [17,18,19]. In our experiment, the graphene edges can either form *ex-situ* and react with ambient molecules or originate in UHV during the DCT process where the graphene edges would consist of C dangling bonds. The edges of the graphene bilayer presented in figure 2 appear pristine and may have formed in UHV, while the edges of the monolayer pieces in figure 4 are decorated with topographic protrusions and were likely generated under ambient conditions. The spectra maps in figure 4 indicate a reduction in graphene bandgap proximal to the bright edge features in the topographs, consistent with gap states arising from these features and similar to the topographic protrusions caused by the SWNT end states observed in previous STM studies [31]. Further experiments will correlate atomic resolution of the edge structures and spatially resolved tunnelling spectroscopy to provide greater insight into the nature of the observed variations in the electronic structure of nanometer-sized graphene.

## 4. Conclusions

In summary, we have developed a method for depositing 2-10 nm sized graphene pieces on a conducting substrate and we have measured that the energy gap of nanometer-sized graphene monolayers scale inversely with the lateral dimensions using tunnelling spectroscopy. Moreover, we can directly quantify the number of graphene layers by measuring the height of the graphene relative to the substrate. Further studies will focus on correlating spatially resolved tunnelling spectroscopy and atomically resolved images of nanometer-sized graphene to determine the dependence of the electronic structure of graphene on lateral size, edge structure, and crystallographic orientation. DCT has previously enabled the deposition of single-walled carbon nanotubes on various metallic [32] and reactive semiconducting [33,34] surfaces including Au(111), Cu(111), Si(100):2x1, GaAs(110), and InAs(110). We anticipate extending STM studies of mechanically exfoliated graphene to the aforementioned substrates to gain deeper insight into the interactions between the nanometer-sized graphene and the supporting substrate.

**Acknowledgements**

This work was supported by the Office of Naval Research under grant number N000140610120 and by the National Science Foundation grant number NSF ECS 04-03489. K. A. R. acknowledges support from a NDSEG fellowship and P. Albrecht and L. Ruppalt for helpful discussions.

*UHV-STM characterization of nanometer-sized graphene on H-Si(100)*

**Figure Captions**

**FIGURE 1. (a)** 100 x 100 nm$^2$ scan of an H-passivated Si(100) surface after depositing graphene. An isolated ≈ 6 nm single layer graphene piece is present at the upper right corner of the image. **(b)** Higher resolution image of the graphene piece from (a). Inset: spatial derivative of a subsequent topographic image where atomic-scale features were observed on the graphene. **(c)** Line contour from the topographic image of the (b) inset derivative image where the position of the contour is designated by the white line. The average spacing between peaks is 2.5 Å in agreement with the distance between nearest neighbor carbon hexagons. **(d)** topographic line contour taken along the gray line in (b) showing the measured graphene monolayer height of ≈ 3.1 Å relative to the Si dimers. Imaging parameters: (a), (b), (b) inset -2 V, 50 pA.

**FIGURE 2. (a)** STM topograph of a 10 x 15 nm$^2$ graphene bilayer. The gray line delineates the position of the line contour shown in (b). Inset: The spatial derivative of the STM topograph highlights the simultaneous atomic-scale resolution of the H-passivated Si dimers and the graphene lattice. **(b)** STM line contour from (a) which displays regions of both single and double layer graphene, denoted by the respective arrows. **(c)** Atomic resolution of bilayer graphene taken from (a). The graphene bilayer displays an STM imaging contrast identical to graphite where 3 of the 6 carbon atoms in hexagons are imaged as protrusions due to the asymmetric stacking of the carbon atoms composing the graphene sheets. A cartoon hexagon with three circles is superimposed to highlight the graphene honeycomb lattice and the three carbon atoms imaged as protrusions. Imaging parameters: (a), (c) -2V, 70 pA.

**FIGURE 3.** *I-V* tunnelling spectra of the graphene monolayer and H-passivated Si from figure 1. Inset: STM topograph where the blue and red dot indicate the position of the graphene and Si tunnelling spectrum, respectively. Imaging parameters: -2V, 50 pA; spectroscopy setpoint: -2V, 1 nA.

**FIGURE 4. (a)** STM topograph and **(b)** topographic spatial derivative of a 5 nm (lower feature) and 2 nm wide (upper feature) single layer graphene pieces. Log(*I*)-*V* spectra plotted as a function of position for the **(c)** 5 nm and **(d)** 2 nm wide graphene monolayers. The red lines in (a) denote the spatial location where the spectra maps were obtained. White dotted lines are used in (c) and (d) to highlight the edges of the graphene monolayers. The bandgap of the graphene monolayers scales inversely with the size of the pieces. The valence ($E_v$) and conductance band ($E_c$) edges are labeled for spectra recorded on the H-passivated Si substrate. Imaging parameters: (a), (b) -2V, 0.1 nA; spectroscopy setpoint: (c), (d) -2 V, 1 nA.

*UHV-STM characterization of nanometer-sized graphene on H-Si(100)*

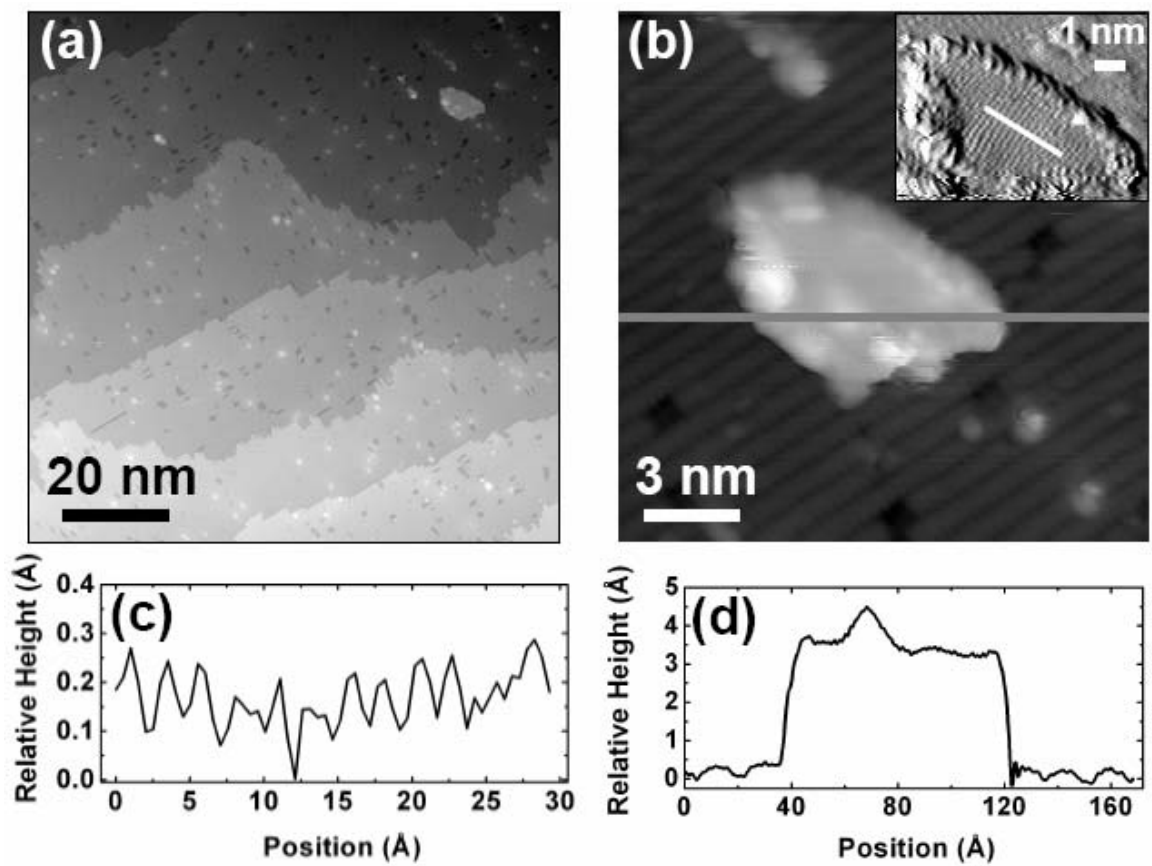

**Figure 1**

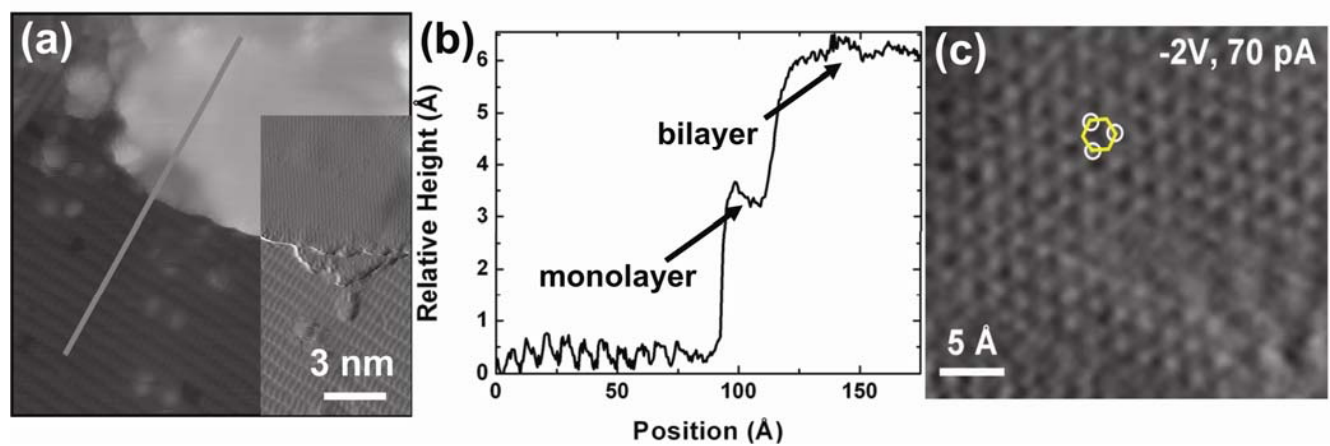

**Figure 2**

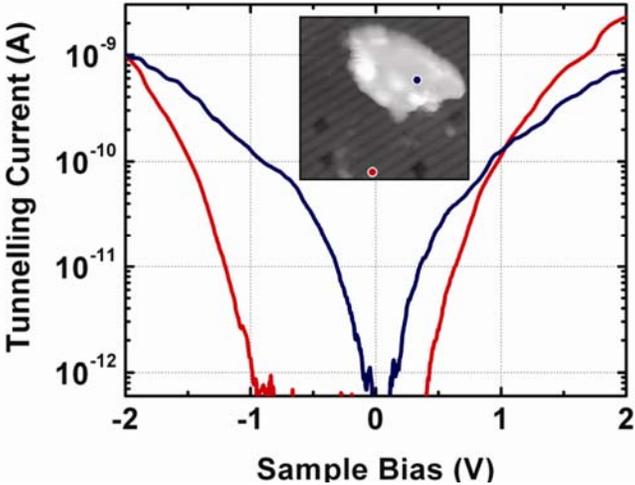

**Figure 3**

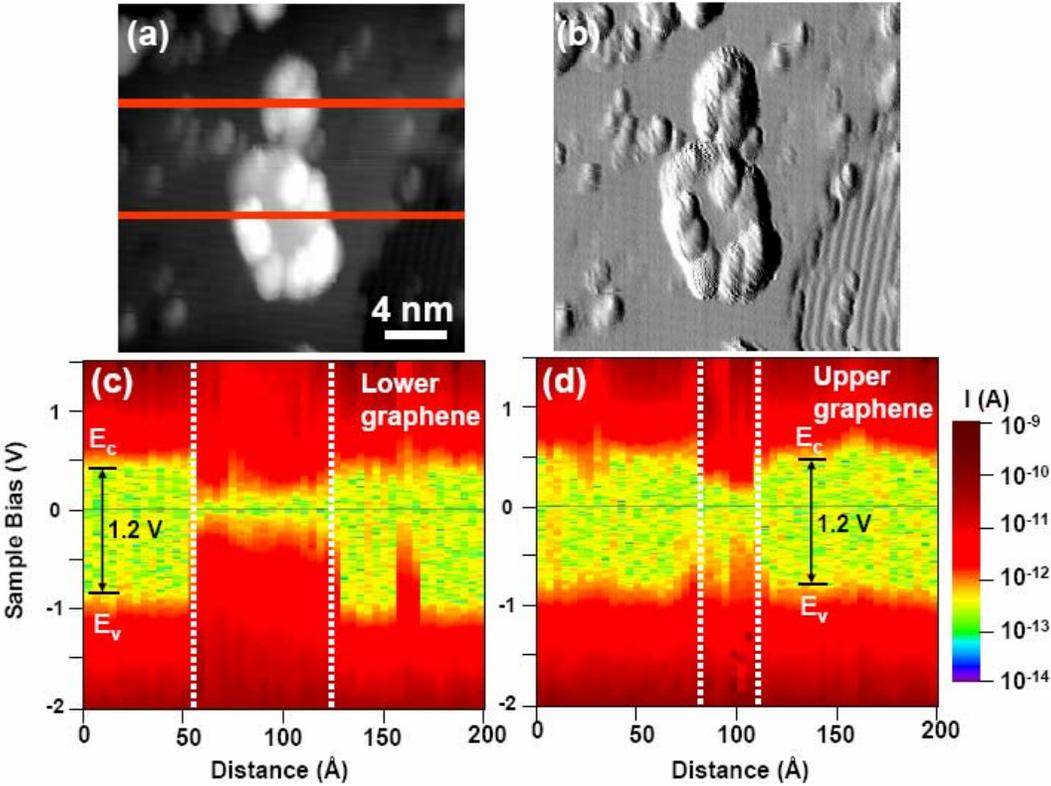

**Figure 4**